\documentclass[useAMS, referee]{mn2e}
\usepackage{graphicx}

\begin{document}
\title{Relationship between eruptions of active-region filaments and associated flares and CMEs\footnote{send
offprint request to: Xiao-Li Yan}}

\pubyear{????} \volume{????} \pagerange{?} \onecolumn
\author[Yan et al. ]
 {X.-L. Yan$^{1,2}$, Z.-Q. Qu $^{1}$, D.-F. Kong$^{3}$\\
        $^1$National Astronomical Observatories/Yunnan Astronomical Observatory, Chinese Academy of Sciences, Kunming,\\ Yunnan 650011, P.R. China yanxl@ynao.ac.cn\\
        $^2$Graduate School of Chinese Academy of Sciences, Zhongguancun, Beijing, P.R. China.\\
        $^3$Jiaxing University, Jiaxing, Zhejiang, P.R.China.\\}

\date{Accepted ????.
      Received ????;
      in original form ???}

\pagerange{\pageref{firstpage}--\pageref{lastpage}} \pubyear{???}

\maketitle \label{firstpage}

\begin{abstract}
To better understand the dynamical process of active-region filament eruptions and associated flares and CMEs,
we carried out a statistical study of 120 events observed by Big Bear Solar Observatory (BBSO), Transition Region and Coronal Explorer (TRACE), and the Extreme-ultraviolet Imaging Telescope (EIT) on board Solar and Heliospheric Observatory (SOHO) from 1998 to 2007.
We combined filament observations with the NOAA's flare reports, Michelson Doppler Imager (MDI) magnetograms, and Large Angle and Spectrometric Coronagraph (LASCO) data, to investigate the relationship between active-region filament eruptions and other solar activities. We found that 115 out of 120 (about 96\%) filament eruptions are associated with flares. 56 out of 105 (about 53\%) filament eruptions are found to be associated with CMEs except for 15 events without corresponding LASCO data. We note the limitation of coronagraphs duo to geometry or sensitivity, leading to many smaller CMEs that are Earth-directed or well out of the plane of sky not being detected by near-Earth spacecraft. Excluding those without corresponding LASCO data, the CME association rate of active-region filament eruptions clearly increases with X-ray flare class from about 32\% for C-class flares to 100\% for X-class flares. We also found that the eruptions of active-region filaments associated with Halo CMEs are often accompanied by large flares (18 out of 20 events; $\geq$ M1.0). About 92\% events (11 out of 12) associated with X-class flare are associated with Halo CMEs. Such a result is due to that the Earth-directed CMEs detected as Halo CMEs are often the larger CMEs and many of the smaller ones are not detected because of the geometry and low intensity. The average speed of the associated CMEs of filament eruptions increases with X-ray flare size from 563.7 km/s for C-class flares to 1506.6 km/s for X-class flares. Excluding the active region located in the area more than 50 degrees from the solar center and 5 without corresponding MDI data, the $\beta$ magnetic field configuration (about 47\%; 36 out of 77) is more likely to form eruptive filaments than other ones and there are 33 filament eruptions associated with magnetic flux cancellation, 42 events associated with magnetic flux emergence, 2 events without variation of magnetic field. The average area of emergence regions is 855.9 square arcseconds. These findings may be instructive to not only in respect to the modeling of active-region filament eruptions but also in predicting flares and CMEs.
\end{abstract}


\section{Introduction}
Solar filaments are formed in magnetic loops that
hold relatively cool (5000-8000K) and dense plasmas ($10^{10}-10^{11}cm^{-3}$) suspended in the hot ($10^6$K) corona
above the solar surface. Filaments and prominences refer to the same physical structures on the Sun,
when they are projected onto the disk (filaments) or extending above the limb (prominences).
Moreover, they are observed as the magnetic flux rope having helical structure
(Rust \& Kumar 1994; Titov \& D$\acute{e}$moulin 1999; Amari et al. 1999; Gibson \& Fan 2006; Gilbert et al. 2007).

It is widely accepted that filament eruptions, flares and Coronal Mass Ejections (CMEs) are different aspects of the same physical
process (Shibata et al 1995; Forbes 2000; Priest \& Forbes 2002).
With the high spatial and temporal resolution observations
by satellite-borne telescopes, more and more active-region filaments are able to be observed (Green et al. 2007; Chifor et al. 2007).
The eruption mechanisms of filaments have been studied by many authors. For instance, it has been found that the eruptions of filaments may be caused by
the kink instability (Sakurai, 1976; Ji et al. 2003; Fan 2003, 2005; Kliem et al. 2004; Rust \& Labonte 2005; Williams et al. 2005;
T$\ddot{o}$r$\ddot{o}$k \& Kliem 2004, 2005; Gibson 2006; Alexander et al. 2006; Liu et al. 2007; Cho et al. 2009) or the torus instability (Schrijver et al.
2008; Aulanier et al. 2010). The occurrence of the two types of instabilities may be caused by magnetic cancellation (Chae et al. 2001; Zhang \& Wang 2001; Jiang et al. 2001; Moon et al. 2004; Sterling et al. 2007; Chifor 2007), the change of the topology of loops overlying the filament (Nagashima et al. 2007), magnetic emergence (Feynman \& Martin 1995; Wang et al. 1999; Romano et al. 2003), and a rapid change of magnetic connectivity in a bundle filament threads (Kim et al. 2001).  Observationally, a number of phenomena are found to be associated filament eruptions which can help to discriminate between models, such as the direction of propagation of the brightenings consistent with the direction of the erupting filament (Tripathi et al. 2006), an apparent increase in the homogenization of the filament mass composition prior to filament eruptions (Kilper et al. 2009), endpoint brightening during the rapid ascent of the filaments (Wang et al. 2009), a bipolar double dimming formed near the two ends of the erupted filament (Jiang et al. 2006; Yang et al. 2008), and compact hard X-ray footpoint sources at the endpoints and a ribbon-like footpoint emission extending along the endpoints during the kink evolution of filaments (Liu \& Alexander 2009).

It is also commonly accepted that the occurrence of CMEs is caused by a loss of
stability or equilibrium of the coronal magnetic field (Low 1993; Forbes 2000; Lin \& van Ballegooijen 2005).
Jing et al. (2004) have investigated the relation between filament eruptions (most of their samples are quiescent filaments),
flares and CMEs by using Big Bear Solar Observatory (BBSO) H$\alpha$ full-disk images (Steinegger et al. 2000).
They found that 56\% of their samples were associated with CMEs except for those with no corresponding LASCO data. We note that faint, Earth-directed CMEs are less likely to be detected and this must have an influence on this figure. Gopalswamy et al. (2003) classified the prominence eruptions into two groups (transverse and radial direction events) and found that 72\% of the prominence eruptions were clearly associated with CMEs by using the Nobeyama Radioheliograph. The studies of Wagner et al. (1981), Simnett \& Harrison (1985), and Harrison et al. (1985, 1986) showed the evidence that the CME onset appeared to precede the flare onset for a few CME events. Through the investigation of 95 CME events from 1984 to 1986, Harrison (1990) found that active regions are the sources of, or related to the source of CMEs. In this paper, we focus on the relationship between the active-region filaments, flares and CMEs observed by BBSO, TRACE, and SOHO/EIT. Although the single active region
events relative to the eruptive filament were investigated by many
authors, establishing a firm quantitative relation between them will require more statistical evidence. Moreover, we can make use of these statistics to make predictions in respect to the occurrence of flares and CMEs.

\section{Observations and Method}
The data used in the paper are as follows:

1. Full-disk H$\alpha$ line-center images from Big Bear Solar
Observatory (BBSO), one station of global high-resolution H$\alpha$ network (Steinegger et al. 2000). The image cadence is 1 minute,
and the pixel size is about 1$^\prime$$^\prime$.

2.The Transition Region and Coronal Explorer (TRACE; Handy et al. 1999) 171{\AA}, 195{\AA}, 1600{\AA} images with 0.$^\prime$$^\prime$5 pixel$^{-1}$ and 1$^\prime$$^\prime$ spatial resolution and temporal resolution of about 1 minute.

3.Full-disk Fe XII 195 {\AA} images from the Extreme-ultraviolet
Imaging Telescope (EIT) on board Solar and Heliospheric Observatory (SOHO; Domingo
et al. 1995) with a cadence of
12 minutes and a pixel size of 2.6$^\prime$$^\prime$ (Delaboudini$\grave{e}$re et al. 1995).

4. Full disk 96 minute line-of-sight magnetograms by Michelson Doppler Imager (MDI; Scherrer et al.
1995) on board SOHO (Domingo
et al. 1995).

5. Geostationary Operational Environmental Satellite (GOES) soft X-ray light curves and NOAA's National Geophysical Data Center soft X-ray flare records ($ftp://ftp.ngdc.noaa.gov\protect\newline/STP/SOLAR\_DATA/SOLAR\_FLARES/$).

6. Large Angle and Spectrometric Coronagraph (LASCO; Brueckner et al. 1995) C2 on board SOHO (Domingo
et al. 1995).

First, we used Full-disk H$\alpha$ images from Big Bear Solar
Observatory (BBSO), the movies of filaments loaded in TRACE homepage Web site ($http://trace.lmsal.com/POD/$), daily observation of TRACE, and Full-disk Fe XII 195 {\AA} images from SOHO/EIT to detect the filament eruptions; Second, during the eruption of the filaments, we firstly identified the flares as sudden increases in intensity of H$\alpha$ or EUV flare loops. Then we examined the GOES soft X-ray flux profiles, NOAA's flare records to identify the flares associated with the eruptions of filaments. Finally, we established that the equivalent position angles of the erupting filaments (see the similar discussion of Harrison \& Sime 1989) are included by the spans of CMEs and the eruptive directions of filaments are consistent with that of CMEs. Besides, the CMEs appear in LASCO C2 coronagraph images within about 2 hours after the filament eruptions. If the conditions were satisfied in respect to the three points mentioned above, we took the CME as being the association of filament eruptions. We coaligned the TRACE images and BBSO H$\alpha$ images with SOHO/MDI magnetograms and EIT images so that we can determine the exact position of the filament eruptions with the active region number. Due to the projection effect, we only took into consideration magnetic configuration of active region within about 50 degrees from the solar center. Meanwhile, we determined magnetic flux emergence and cancellation from a time sequences of MDI magnetograms with a 300$^\prime$$^\prime$ $\times$ 300$^\prime$$^\prime$ FOV (for super active region, we adopted 500$^\prime$$^\prime$ $\times$ 500$^\prime$$^\prime$ FOV ) obtained at least 12 hours prior to the filament eruption. We evaluated the magnetic flux emergence and cancellation from the magnetic field in the vicinity of the filament (Feynman \& Martin 1995).

\section{Results of filament eruption associations}
\subsection{Distribution of samples}
We found 120 active-region filament eruptions for which we obtained the complete observation of the eruption. Tables 1, 2 present a list of 120 active-region filament eruptions and summarize their relation to flares, CMEs, magnetic configuration, and the change of magnetic field. The columns contain observation time, active region number, the time and position of the filament eruption onset, associated flares and CMEs, CME speed, magnetic configuration, and magnetic flux emergence or cancellation. Some of these filaments were mentioned by Jing et al. (2004), Green et al. (2007), Williams et al. (2005), Rust \& Labonte (2005), Schrijver et al. (2008), Yurchyshyn et al. (2001) and other researchers. Figure 1 shows the latitude distribution of the active-region filaments in our samples on the solar disk from 1998 to 2007. It is notable, and expected, that this figure resembles the typical butterfly diagram following the solar cycle. The distribution
of active-region filaments yearly is plotted in Figure 2 and, again, this peaks with the solar cycle.

\subsection{Relation between active-region filaments, flares, and CMEs}
Before examining the statistical associations, we took one examples as a means of illustrating the relationship of active-region filaments to flares, CMEs. Figure 3 shows the filament eruption observed by SOHO/EIT in active region 9163 on 2000 September 12 (fig. 3a). The filament reached instability at 10:12UT. After the eruption of this filament, an M1.0 flare occurred in this region (fig. 3b). Additionally, a Halo CME was found to be associated with the eruption of this filament (fig. 3c, d). Before the eruption of the filament, we found that the magnetic flux emergence (as is shown in the circled area) appeared in the vicinity of the eruptive filament (fig. 3e, f).

Figure 4 shows the heliographic latitude of flares and CMEs, which were associated with active-region filament eruptions. We used pluses to denote filament eruptions associated with flares. Asterisks denote the occurrence of both flares and CMEs. The triangles indicate the eruptions, which are associated with neither flares nor CMEs. The squares denote the fact that CMEs only took place respectively.

In our samples, 115 out of 120 (96\%) of filament eruptions are found to be associated with flares. Jing et al. (2004) investigated 21 active-region filament and 85 quiescent filament eruptions. They found that 95\% of active-region filament and 27\% of quiescent filament eruptions are associated with flares. We obtained the similar results as Jing et al. (2004) on active-region filament by using more samples. Compared with their results on quiescent filaments, the active-region filaments have higher flare productivity.

Except for the 15 events without corresponding LASCO data, 56 out of 105 (about 53\%) active-region filament eruptions are associated with CMEs. Jing et al. (2004) obtained 54\% CME association rate on quiescent filaments. Gopalswamy et al. (2003) and Gilbert et al. (2000) found that 73\% and 94\% of the prominence eruptions are associated with CMEs. Their results are much higher than what we and Jing et al. (2004) obtained. Note that most of our samples as well as Jing et al.'s were taken from an area located in the solar disk. Filaments are a component of the CME eruptions. However, the weaker Earth-directed CMEs and the events well out of the plane of the sky cannot be as easily detected by the coronagraphs as events near the plane of the sky. Thus, our results must be considered in the light of observational issues (Yashiro et al. 2005).

Out of 13 events in which one filament eruption have no corresponding CME data, the eruptions of filaments associated with X-class flare are all associated with CMEs. Since we limited our sample (only 12 events), this result mentioned above need more events to confirm. Out of 43 events in which 7 filament eruptions have no corresponding CME data, 25 out of 36 (about 69\%) events associated with M-class flare are associated with CMEs. Out of 52 events in which 5 filament eruptions have no corresponding CME data, 15 out of 47 (32\%) events associated with C-class flare are associated with CMEs. Figure 5 shows the distribution of associated CME speed by associated flare class. The average speeds of CMEs associated with C-class, M-class, and X-class flares are 563.7, 770.3, and 1506.6 km/s respectively. 54 out of 105 (about 51\%) active-region filaments produce both flares and CMEs. Simultaneous occurrence rate of associated flares and CMEs obtained by us is higher than that obtained by Jing et al. (2004) (9 out of 19; about 47\%).

\subsection{The magnetic configuration of active regions with filaments}
If the active region is located far away from solar disk center, the projection effect affects the judgement of magnetic configuration. Excluding the active regions located beyond 50 degrees from the solar center, the numbers of different magnetic configurations (Hale \& Nicholson 1938) are shown in the figure 6. The $\beta$ magnetic field configuration (about 47\%; 36 out of 77) is more likely to form eruptive filaments than other magnetic configurations. Among these examples, 20 Halo CMEs are found. Excluding four active regions located in the area more than 50 degrees from the solar center, the active regions with $\beta$$\delta$$\gamma$ (7 out of 16) and $\beta$ (8 out of 16) magnetic field configuration are more likely to produce Halo CMEs than other active regions.

\subsection{The change of magnetic field before eruptions of filaments}
For the same reason mentioned above, we excluded the active regions beyond 50 degrees from the solar center and 5 ones lacking MDI data. After the differential rotation correction, we filmed a sequence of MDI magnetograms in order to examine the change of magnetic fields. The magnetic flux emergence and cancellation appeared in the vicinity of an filament and at least 12 hours prior to the filament eruption (Jing et al. 2004). We found that there are 33 filament eruptions associated with magnetic flux cancellation, 42 events associated with magnetic flux emergence, and 2 with neither of these. The size of the emergence region of the magnetic flux is from 188 $arcsec^2$ to 1924 $arcsec^2$ (fig. 7). The average area is 855.9 $arcsec^2$.

\subsection{Halo CME associations}
A CME with angular width of 360 degree is defined as a Halo CME. In our samples, 20 Halo CMEs are found associated with eruptions of active-region filaments. The eruptions of active-region filaments associated with Halo CMEs are often accompanied by large flares (18 out of 20; $\geq$M1.0). Out of 20 events, about a half events (11 out of 20; 55\%) are associated with X-class flares and 7 out of 20 events (35\%) are associated with M-class flares. Out of 13 events in which one filament eruption have no corresponding CME data, 11 out of 12 (about 92\%) events associated with X-class flare are associated with Halo CMEs. The Earth-directed CMEs detected as Halo CMEs are almost certainly the larger event; smaller CMEs so far from the plane of the sky are less well detected by coronagraphs. That is to say, the smaller CMEs that are not detected by the coronagraphs may be also associated with large flares.

\section{Conclusion and Discussion}
In this paper, we focused on the statistical relationship between eruptions of active-region filaments and associated flares and CMEs. We identified 120 active-region filament eruptions observed by BBSO, TRACE, and SOHO/EIT from 1998 to 2007. The main results are summarized as follows:

(1) Of the 120 events, 96\% filament eruptions are found to be associated with flares;

(2) Excluding 15 events without corresponding LASCO data, 56 out of 105 (about 53\%) filament eruptions are found to be associated with CMEs;

(3) Out of 13 events in which one event has no corresponding LASCO data, the eruptions of filaments associated with X-class flare are all associated with CMEs.

(4) About 92\% (11 out of 12) events associated with X-class flare are associated with Halo CMEs.

(5) Out of 43 events in which seven events have no corresponding LASCO data, 25 out of 36 (about 69\%) filament eruptions associated with M-class flare are associated with CMEs.

(6) Out of 52 events in which five events have no corresponding LASCO data, 15 out of 47 (about 32\%) filament eruptions associated with C-class flare are associated with CMEs.

(7) The average speed of the associated CMEs of filament eruptions increases with X-ray flare size from 563.7 km/s for C-class to 1506.6 km/s for X-class flares.

(8) 54 out of 105 (about 51\%) active-region filaments produce both flares and CMEs.

(9) The eruptions of active-region filaments associated with Halo CMEs are often accompanied by large flares (18 out of 20; $\geq$M1.0).

(10)Except for the active region located in the area more than 50 degrees from the solar center and 5 without corresponding MDI data, we found that the $\beta$ magnetic field configuration (about 47\%; 36 out of 77) is more likely to form eruptive filaments than other ones and there are 33 filament eruptions associated with magnetic flux cancellation, 42 events associated with magnetic flux emergence, 2 events without variation of magnetic field. The average area of emergence regions is 855.9 square arcseconds.

Feynman \& Martin (1995) and Jing et al. (2004) reported that 42\% and 54\% of the eruptions of quiescent filaments are associated with CMEs respectively.
Compared with our results, it can be see that the CME association rate of active-region filament eruptions is nearly the same as that of quiescent filaments. We also found that active-region filaments have higher flare production than quiescent ones by using more samples. Gopalswamy et al. (2003) and Gilbert et al. (2000) reported much higher (73\% and 94\%) associated CMEs productions of the prominence eruptions than what we obtained (about 53\%). Note that most of our samples as well as Jing et al.'s were located in the solar disk. These results might be caused by that the slow and narrow CMEs may not be visible when CMEs originate from the disk center (Yashiro et al. 2005). Without a doubt, the typical CME is the classical three-part structure with a bright outer envelope, a dark inner void, a bright filamentary core. The filament to many researchers is a part of the CME structure. Even if one do not see the internal filamentary structure for some events, in coronagraph data one never see an erupting filament without what one would call a CME. Normally, it may be useful to talk about how many CMEs contain filaments. In this paper, we focused on how many filament eruptions are associated with CMEs. Indeed, our results are affected by the detection capability of the coronagraphs. The coronagraphs did not detect the CME due to geometry or sensitivity -i.e., weaker Earth-directed CMEs, or the events well out of the plane of the sky may not be detected by the coronagraphs.

And interestingly, we found that the eruptions of active-region filaments associated with X-class flare are all associated with CMEs except for one event without LASCO data. We need make use of forthcoming data to confirm it. This result is very useful to predict the occurrence of CME. In addition, the eruptions of active-region filaments associated with Halo CMEs are often accompanied by large flares (18 out of 20; $\geq$M1.0). The point that merits our particular attention is that about 92\% events (11 out of 12) associated with X-class flare are associated with Halo CMEs. The coronagraphs on board near-Earth spacecraft can easily detect Halo CMEs that have large angular width. However, the many smaller CMEs can not be detected duo to the geometry and low intensity. Our result shows that the Halo CMEs are often associated with large flares and we do not rule out that the smaller CMEs that are not detected by the coronagraphs may be also associated with large flares.

The instability of the magnetic loop is a hot topic in solar
physics. The total twist in a given magnetic loop is a
major factor in MHD stability. Some of sudden
magnetic energy release such as flares (Hood
\& Priest 1979), and coronal mass ejections are
probably due to the kink instability of filaments(Forbes \& Isenberg 1991; Lin et al. 1998; Fisher et al. 1999; Liu et al. 2003; Leka et al. 2005; Zhou et al. 2006), the torus instability of a flux rope (Roussev et al. 2003; Kliem \& T$\ddot{o}$r$\ddot{o}$k 2006), or the reconnection
between closed field lines in the streamer belt and adjacent, open field lines (Harrison et al. 2009). In particular, the eruption of the transient sigmoid (Canfield et al. 1999; Pevtsov 2002; Low \& Berger 2003) or a
filament with apex rotation is a more often observed phenomenon
(T$\ddot{o}$r$\ddot{o}$k \& Kliem 2005; Green et al. 2007). Ji et al. (2003) observed a failed eruption of a filament and suggested
that at least a reconnection very low in the corona (possibly above the filament)
and open (opening ) field above that point are required for successful eruption. While, some authors suggested that
reconnection beneath the erupting footpoint may destabilise the filaments or
prominences and cause the eruptions (Moon et al. 2004; Chifor et al. 2006; Nagashima et al. 2007, Sterling et al. 2007). The magnetic flux emergence and cancellation were found before the eruptions of a portion of filament eruptions in our samples. Consequently, we think that magnetic flux emergence and cancellation play an important role in triggering eruptions of active-region filaments in some events. Undoubtedly, there are other factors that trigger filament eruptions. Zhang et al. (2004) suggested that CMEs large-scale acceleration
and flare particle acceleration are driven by the same
physical process or by multiple processes that are physically
coupled in the corona. The topics relating to the way in which the instability of twisted loops comes into being and to the conditions that can cause their eruptions deserve us to further investigation.

\section*{Acknowledgments}
The authors thank the referee for the careful reading of the manuscript and constructive
comments that improved the original version. The authors thank the BBSO, TRACE, SOHO, NOAA and GOES consortia for their
data. SOHO is a project of international cooperation between ESA and
NASA. Global High Resolution Ha Network is operated by the Space Weather Research Lab, New Jersey Institute of Technology. This work is supported by the National Science Foundation of China (NSFC) under grant numbers 10903027, 10673031, 10943002 and 40636031, Yunnan Science Foundation of China under grant number 2009CD120, and the
National Basic Research Program of China 973 under grant number
G2006CB806301.





\clearpage

\begin{figure}
   \centering
   \includegraphics[width=10cm]{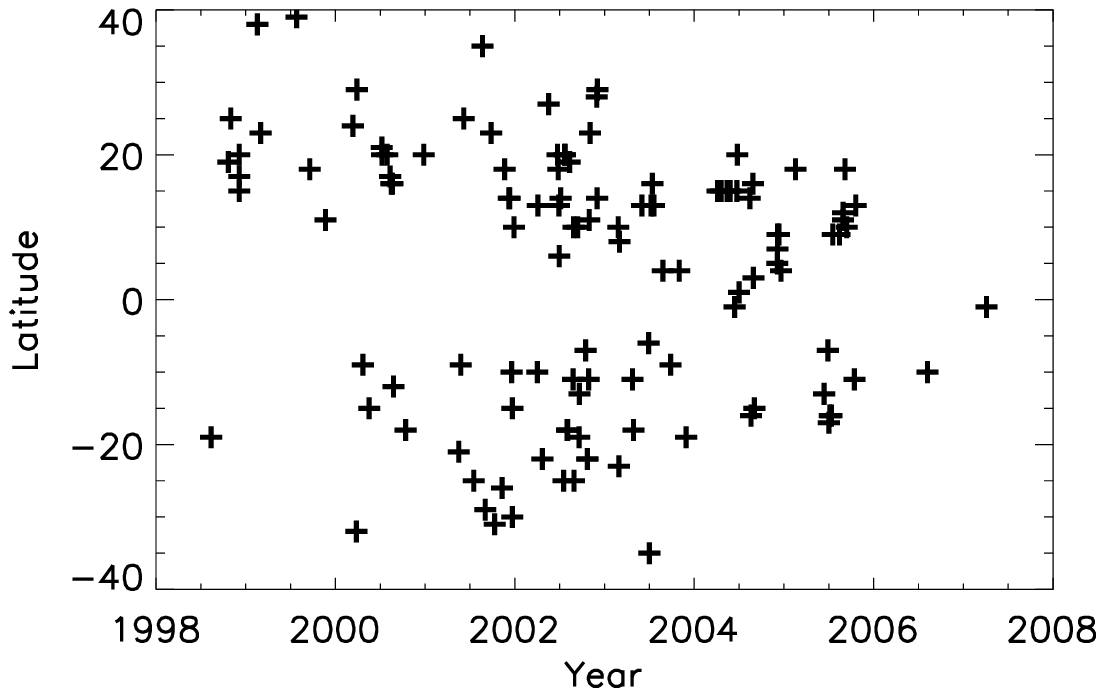}
\caption{The distribution of active-region filaments on the solar disk from 1998 to 2007.}
        \label{}
  \end{figure}

 \begin{figure}
   \centering
   \includegraphics[width=10cm]{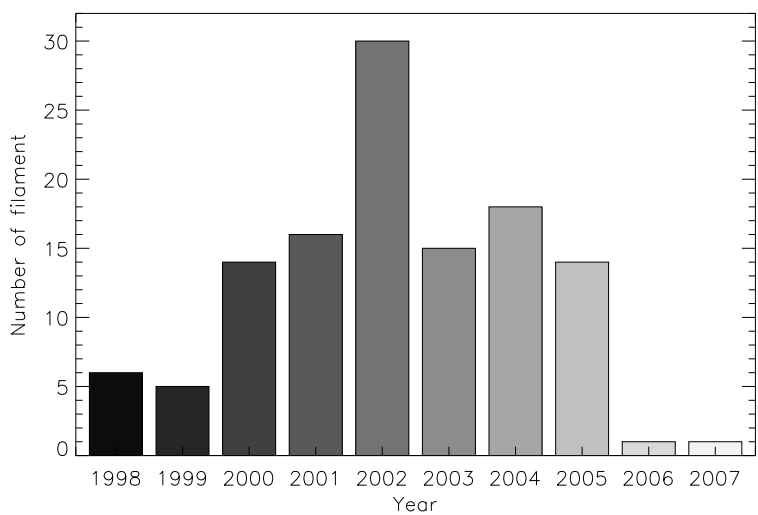}
\caption{The distribution of active-region filaments yearly in our samples from 1998 to 2007.}
        \label{}
  \end{figure}

\begin{figure}
   \centering
   \includegraphics[width=5cm]{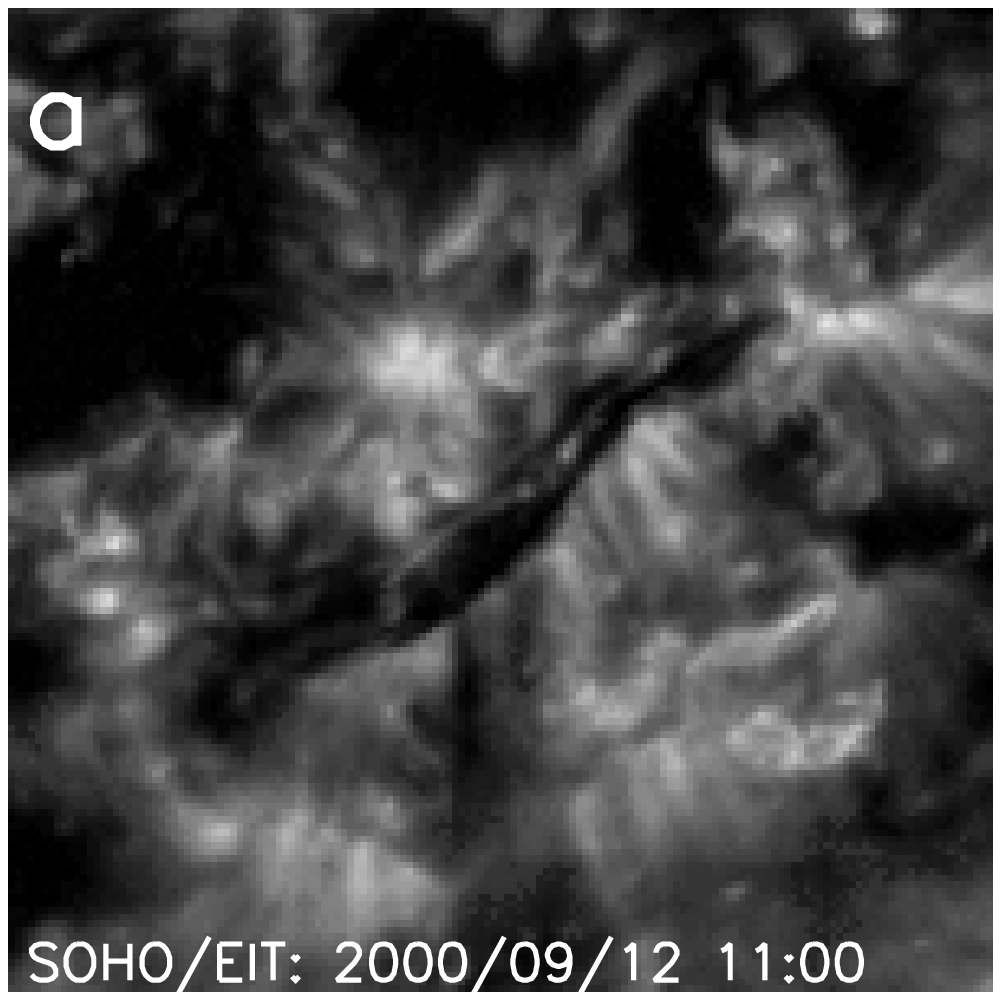}%
   \includegraphics[width=5cm]{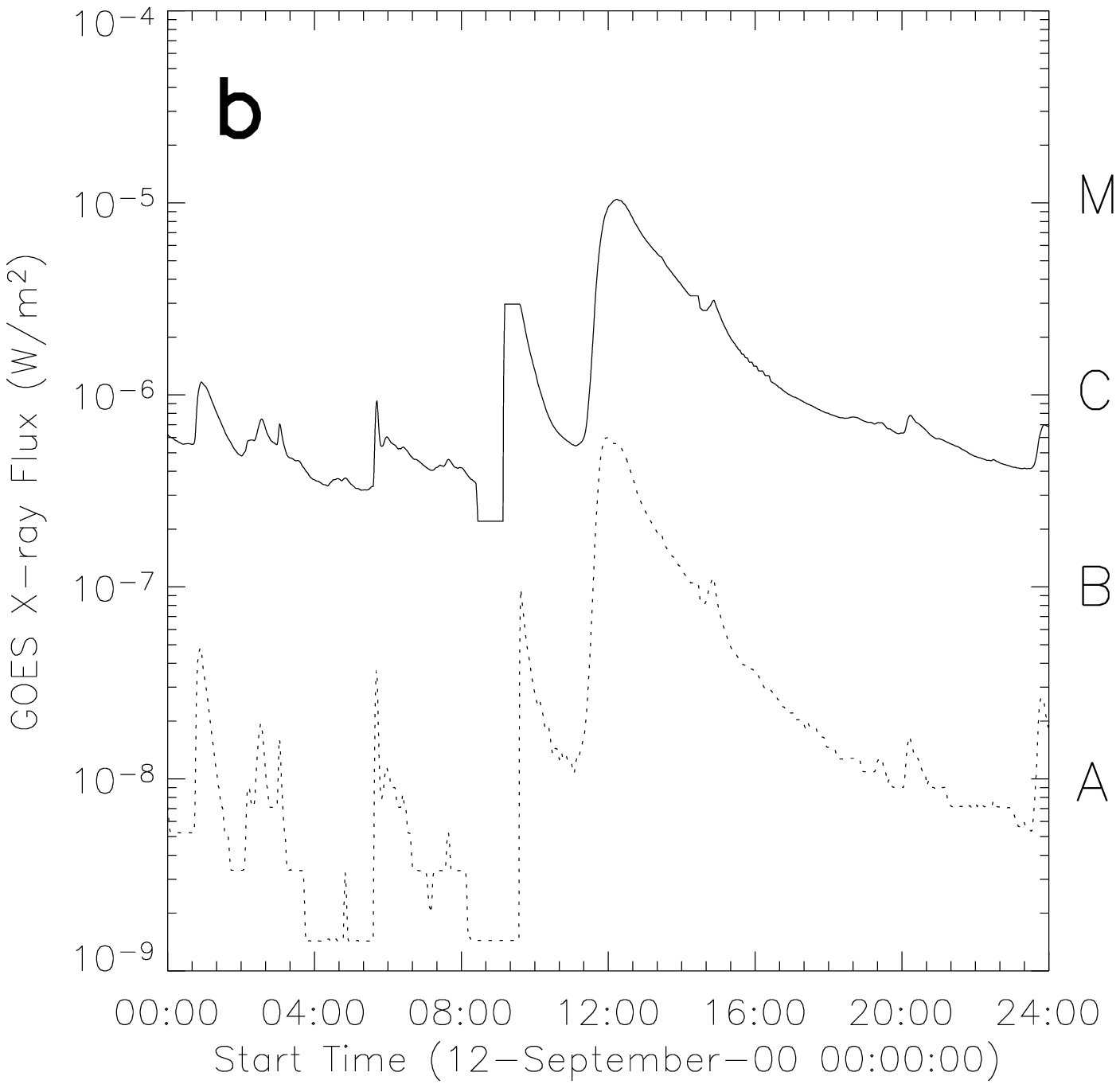}\\
   \includegraphics[width=5cm]{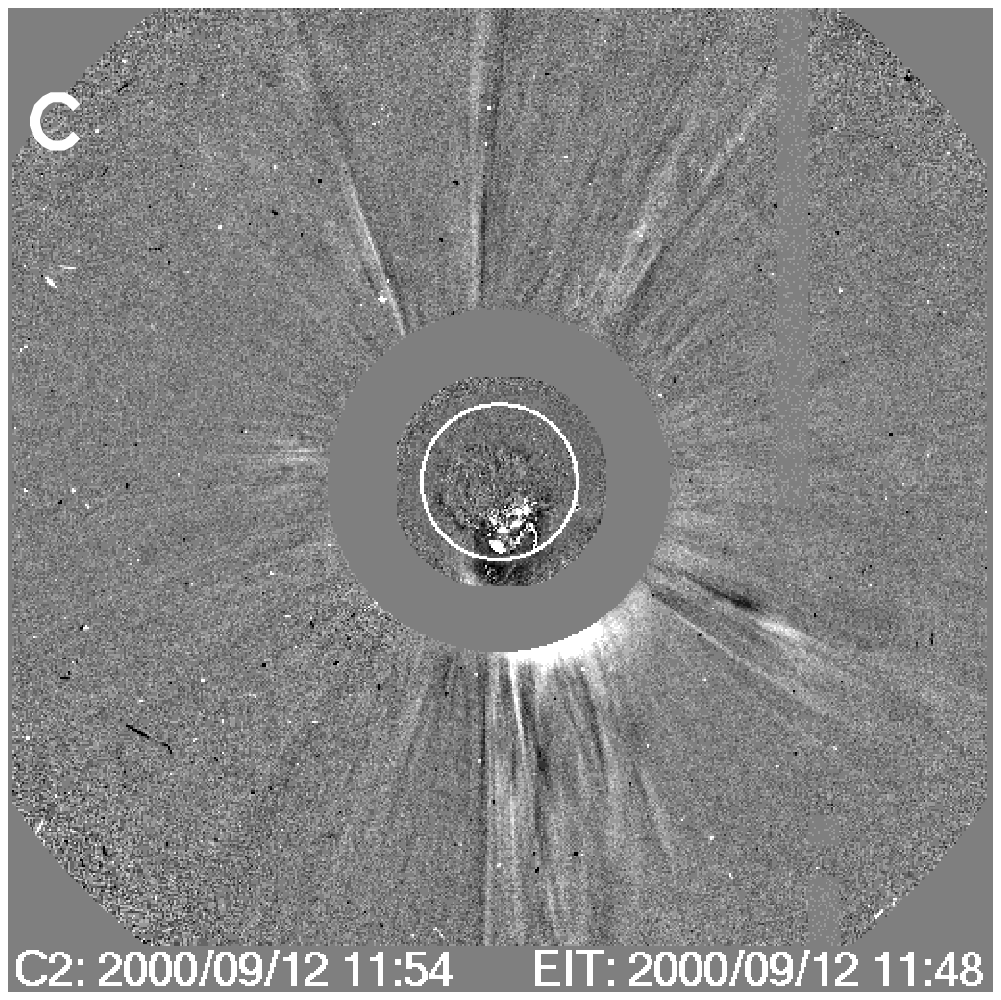}%
   \includegraphics[width=5cm]{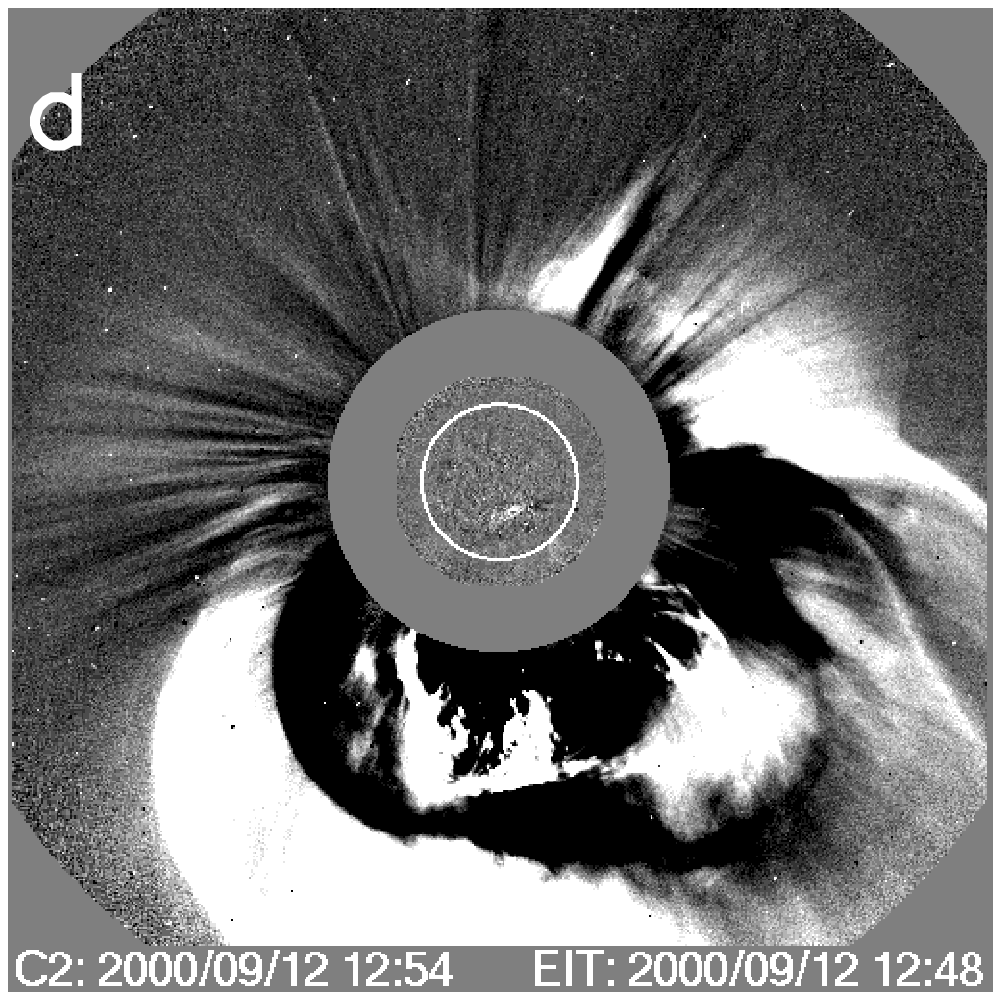}\\
   \includegraphics[width=5cm]{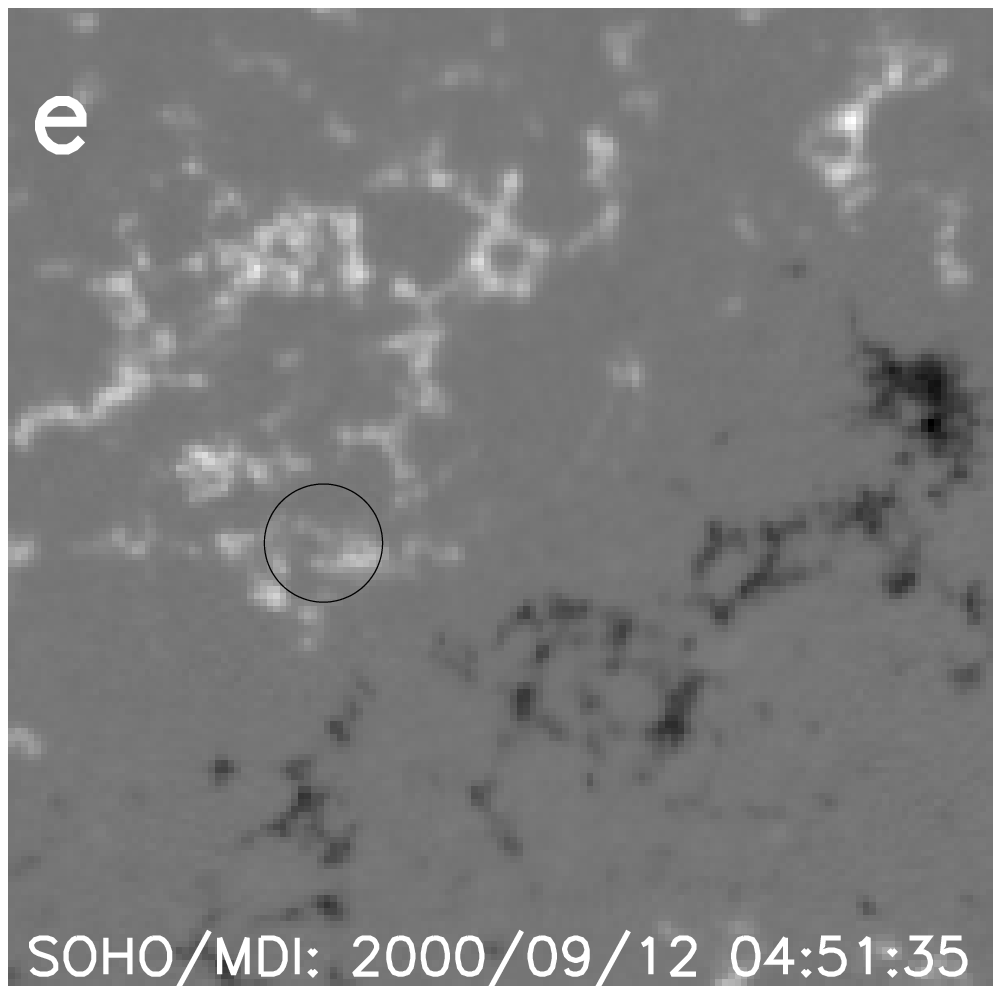}%
   \includegraphics[width=5cm]{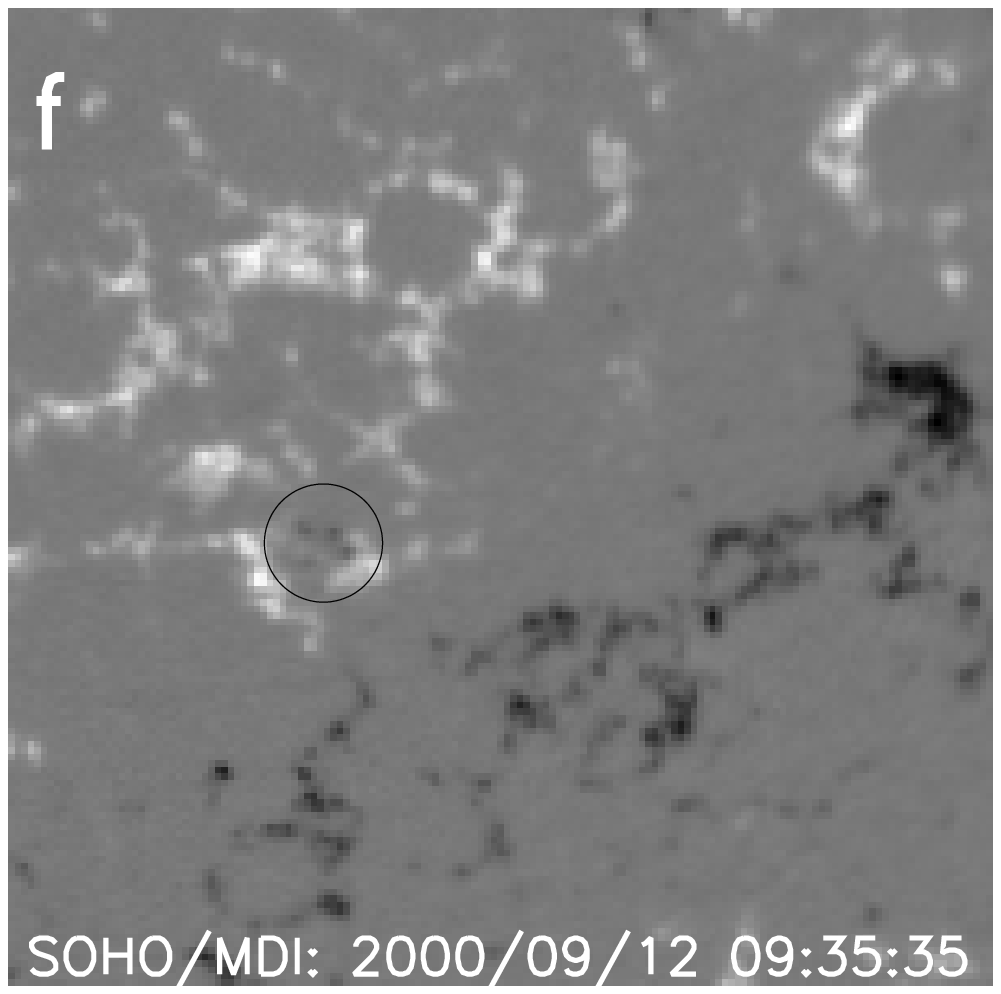}\\
\caption{a: A filament in Active Region 9163 observed by SOHO/EIT. b: Evolution of GOES soft X-ray emission (Solid line: 1-8
\AA; Dashed line: 0.5-4 \AA ) for the M6.3 flare on 2000 September 12. c and d: SOHO/LASCO running difference images of the white-light CME after eruption of the filament. e and f: The SOHO/MDI magnetograms of active region 9163. The circles indicate the position of magnetic flux emergence.}
        \label{}
  \end{figure}

  \begin{figure}
   \centering
   \includegraphics[width=10cm]{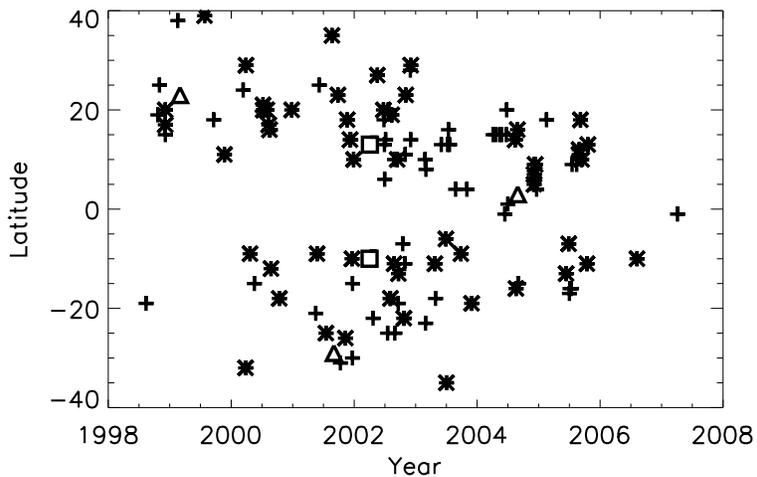}
\caption{Latitude distribution of eruptive filaments and their overall relation to flares and CMEs. Pluses denote filament eruptions, which were associated with flares. Asterisks refer to filament eruption associated with both flares and CMEs, triangles to neither flares nor CMEs related, squares to only CMEs.}
        \label{}
  \end{figure}

\begin{figure}
   \centering
   \includegraphics[width=10cm]{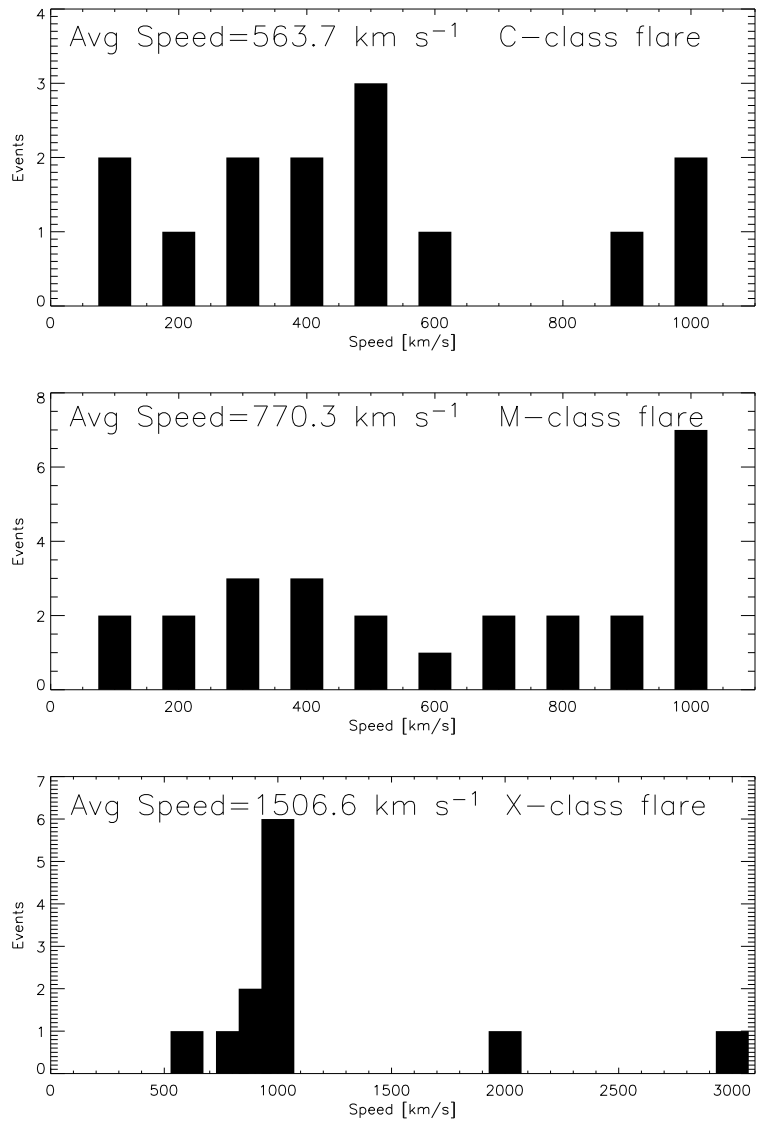}
\caption{Distribution of associated CME speed by associated flare size respectively.}
        \label{}
  \end{figure}

\begin{figure}
   \centering
   \includegraphics[width=10cm]{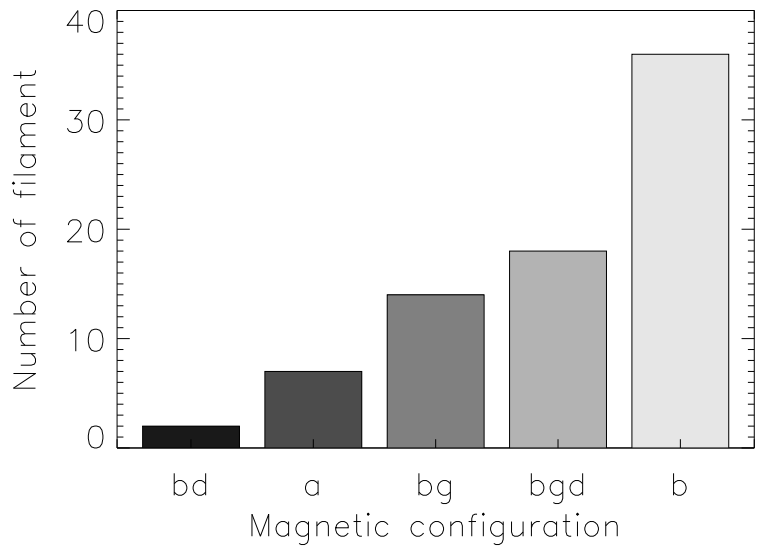}
\caption{The numbers of active-region filaments with different magnetic configurations. ``bd",``a", ``b", ``bg", and ``bgd" indicate $\beta$$\delta$, $\alpha$, $\beta$, $\beta$$\gamma$, $\beta$$\gamma$$\delta$ respectively.}
        \label{}
  \end{figure}

\begin{figure}
   \centering
   \includegraphics[width=10cm]{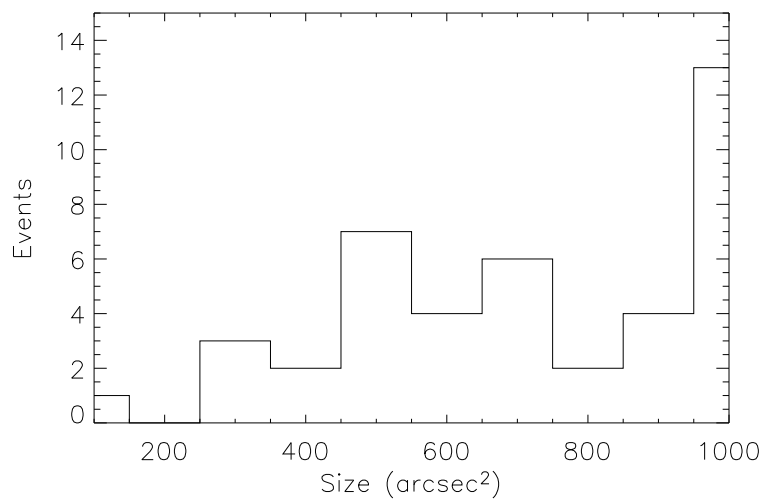}
\caption{The numbers of events with the different sizes of the emergence regions of the magnetic flux.}
        \label{}
  \end{figure}

\clearpage
\begin{table*}{}
\begin{center}
TABLE 1 \\
A list of active-region filament eruptions and their associated solar activities. ``AR" indicates the active region number.
``Time" and ``Location" indicate the time and position of the filament eruption onset. ``CME Speed" indicates linear
Speed of CME. ``Hale" indicates the magnetic configuration of the active region. ``MFE" and ``MFC" indicate the magnetic flux emergence and cancellation. ``No Data" in the sixth and ninth columns denotes no LASCO observation data covering the process of the filament eruptions and no magnetograms data prior to the filament eruptions. ``No" in the ninth columns denotes no obvious change of magnetic fields before filament eruption. A CME with angular width of 360 degree is defined as a Halo CME.\\
\vspace{0.5cm}
\begin{tabular}{llllllllll}
\hline\hline \label{tbl-1}
Date & AR & Time & Location  & Flares& CMEs  &CME Speed (km $s^{-1}$) &Hale & MFE or MFC \\
\hline
1998 Jul 11 & 8260 &04:21 & S19W80      &  C1.7       &  No Data &$\cdot\cdot\cdot$  &  $\cdot\cdot\cdot$  & $\cdot\cdot\cdot$  \\
1998 Sep 20& 8340 & 02:30 & N19E37    & M1.8  & No Data &$\cdot\cdot\cdot$& $\cdot\cdot\cdot$& No data\\
1998 Sep 30 &8340 &11:54& N25W90 &  M2.8&    No Data&$\cdot\cdot\cdot$& $\cdot\cdot\cdot$   &$\cdot\cdot\cdot$\\
1998 Nov 4 & 8375 &03:14& N17E01 & C5.2 & CME& 102& $\beta$$\gamma$&MFE\\
1998 Nov 4 & 8375 &07:13 &N20E01&C1.6 & Halo &523& $\beta$$\gamma$& MFC\\
1998 Nov 4 & 8375 &22:02 &N15W17 & C2.4 & $\cdot\cdot\cdot$ &$\cdot\cdot\cdot$ & $\beta$$\gamma$ &MFE\\
1999 Jan 17 &8440 & 17:32  &N38E16 & C2.5&  No data&$\cdot\cdot\cdot$& $\cdot\cdot\cdot$& No data\\
1999 May 31&8559 &22:06&N23W19&$\cdot\cdot\cdot$&$\cdot\cdot\cdot$&$\cdot\cdot\cdot$& $\beta$& MFC\\
1999 Jun 24 &8595 &12:48 &N39W08 & C4.1 & Halo &974.9& $\beta$&MFC\\
1999 Aug 17&8668 &09:18&N18E32&C2.6&$\cdot\cdot\cdot$&$\cdot\cdot\cdot$& $\beta$&MFE\\
1999 Oct 20&8731 &05:51&N11W46&M1.7&CME&326.5& $\beta$$\gamma$ &MFC\\
2000 Feb 10 & 8858&01:04 &N24E01 & C7.3&$\cdot\cdot\cdot$ &$\cdot\cdot\cdot$& $\beta$&MFC \\
2000 Feb 17 & 8872 &20:04&S32E07 & M1.3 & Halo &727.7& $\beta$& MFC\\
2000 Feb 26 & 8889 &21:47& N29E53 & M1.0 & CME &667.7& $\cdot\cdot\cdot$ &$\cdot\cdot\cdot$\\
2000 Mar 20 & 8921 &08:12 & S09E90 & M2.2 & CME &1225.7& $\cdot\cdot\cdot$&$\cdot\cdot\cdot$ \\
2000 Apr 16& 8948 &03:51 & S15W80& C4.2 & $\cdot\cdot\cdot$ &$\cdot\cdot\cdot$& $\cdot\cdot\cdot$ &$\cdot\cdot\cdot$ \\
2000 Jun 6 & 9026& 13:30& N24E18 & M7.1 & CME &357.7& $\beta$$\gamma$$\delta$& MFC\\
2000 Jun 7 & 9026 & 15:30& N21E02 & X1.2 & Halo &842& $\beta$$\gamma$$\delta$ & MFC\\
2000 Jun 28 & 9051 &18:31& N20W90 & C3.7 & CME&1198&  $\cdot\cdot\cdot$& $\cdot\cdot\cdot$\\
2000 Jul 10 & 9070 &19:40& N17W40 & M1.9 & CME&425.8& $\beta$$\gamma$& MFC\\
2000 Jul 14 &9077&09:30& N16E01& X5.7 & Halo &1674.2&  $\beta$$\gamma$$\delta$ & MFC\\
2000 Jul 19 & 9077 & 14:08& N16W70& C3.4 & No data &$\cdot\cdot\cdot$& $\cdot\cdot\cdot$&$\cdot\cdot\cdot$ \\
2000 Jul 23 & 9091 & 04:24& S12W06 & Optical  & CME&630.8 & $\alpha$& MFE \\
2000 Sep 12&9163 &11:03& S18W10&M1.0 & Halo &1550.1& $\beta$&MFE\\
2000 Nov 24 & 9236 &17:20& N20W01 & X1.8 & Halo &1005.5& $\beta$& MFE \\
2001 Apr 23 & 9431 & 12:03& S09W28 & C2.8 & CME&529.7& $\beta$&MFE \\
2001 Apr 15 & 9415 &21:58& S21W80 & C5.1 & $\cdot\cdot\cdot$&$\cdot\cdot\cdot$ & $\cdot\cdot\cdot$&$\cdot\cdot\cdot$  \\
2001 May 5 & 9445 &18:01& N25W11 & C6.3 & $\cdot\cdot\cdot$&$\cdot\cdot\cdot$ & $\beta$$\gamma$&  MFE\\
2001 Jun 15 & 9502 & 09:58& S25E39 & M6.3 & CME &1090.4&  $\beta$& MFC\\
2001 Jul 20& 9538 &03:36& N35W27& B7.6  & CME &193.4& $\alpha$& MFE\\
2001 Aug 1 & 9557 &20:28& S29W14 &$\cdot\cdot\cdot$ & No data& $\cdot\cdot\cdot$ &$\beta$&MFC \\
2001 Aug 25 & 9591 &16:32& S23E45& X5.3  & Halo&1432.8 & $\beta$$\gamma$$\delta$ & MFE\\
2001 Sep 9 & 9608 &20:38& S31E29 & M9.5 & $\cdot\cdot\cdot$ &$\cdot\cdot\cdot$& $\beta$$\gamma$& MFE\\
2001 Oct 9 & 9653 & 10:48& S26E03& M1.4   & Halo &973 &$\beta$& No \\
2001 Oct 19 & 9661 & 16:15& N18W40& X1.6  & Halo &901& $\beta$$\gamma$$\delta$ &MFE \\
2001 Oct 28 & 9682 &16:43& N15E29& M1.4  & CME&279.4 & $\beta$$\gamma$$\delta$ & MFE\\
2001 Nov 1 & 9682 & 11:32 &N14W30 & M3.3 &$\cdot\cdot\cdot$& $\cdot\cdot\cdot$ & $\beta$$\gamma$$\delta$ &MFC\\
2001 Nov 8 & 9690 & 15:45& S30E23 & M4.2 & No data&$\cdot\cdot\cdot$ & $\beta$$\gamma$ & MFE\\
2001 Nov 9& 9687 & 18:20& S15W43 & M1.9 & No data&$\cdot\cdot\cdot$& $\beta$$\gamma$$\delta$ &MFC \\
2001 Nov 17 & 9704 & 04:04& S10E47 & M2.8 & Halo &1379 & $\beta$ & MFE\\
2001 Dec 1 & 9724 &17:39 & N10E80 & M1.8 & CME  &736& $\cdot\cdot\cdot$& $\cdot\cdot\cdot$\\
2002 Mar 2 & 9856 & 13:25& S10E90&$\cdot\cdot\cdot$& CME&1131.2 & $\cdot\cdot\cdot$&$\cdot\cdot\cdot$\\
2002 Mar 21 & 9871 &18:05 & S22W29 & C1.3 & $\cdot\cdot\cdot$&$\cdot\cdot\cdot$& $\beta$& MFC\\
2002 Apr 9 & 9885 & 07:15&N13W90&$\cdot\cdot\cdot$ & CME &260.3& $\cdot\cdot\cdot$& $\cdot\cdot\cdot$\\
2002 Apr 16 &9893 & 12:53& N27W80 & M2.5 & CME&166 &$\cdot\cdot\cdot$ & $\cdot\cdot\cdot$\\
2002 May 21 & 9960 & 20:16& N20E40 &M1.5 & CME &853.3 &$\beta$ & MFE \\
2002 May 24 & 9962 &20:26 & N18E14 & Optical & $\cdot\cdot\cdot$ &$\cdot\cdot\cdot$& $\beta$ & MFC \\
2002 May 27 & 9957 &18:00& N13W75 & M2.0 &  $\cdot\cdot\cdot$ &$\cdot\cdot\cdot$&$\cdot\cdot\cdot$ & $\cdot\cdot\cdot$\\
2002 May 28 & 9957 &16:19& N06W90 & C3.6 &  $\cdot\cdot\cdot$ &$\cdot\cdot\cdot$&$\cdot\cdot\cdot$ &$\cdot\cdot\cdot$ \\
2002 Jun 4 & 9974 &17:47 & N14W23 & C1.0 & No data&$\cdot\cdot\cdot$ & $\alpha$& MFE\\
2002 Jun 16 & 9991 &21:13&S25W45& C1.2 & $\cdot\cdot\cdot$&$\cdot\cdot\cdot$ & $\beta$& MFE \\
2002 Jun 17 & 10001 &22:53& N20E40 & Optical & $\cdot\cdot\cdot$&$\cdot\cdot\cdot$ & $\beta$& MFC \\
2002 Jun 19 & 10000&20:05& N20W15 & Optical &$\cdot\cdot\cdot$& $\cdot\cdot\cdot$ &$\beta$& MFE \\
2002 Jul 1 & 10016 &20:32& S18W09 & C1.0  & $\cdot\cdot\cdot$ &$\cdot\cdot\cdot$&$\beta$& MFE \\
2002 Jul 4 &10019 &16:09& S18E06 & C3.4 & CME &343.1& $\beta$$\gamma$& MFE \\
2002 Jul 15 & 10030 &20:04& N19E01 & X3.0 & Halo &1151& $\beta$$\gamma$$\delta$& MFC \\
2002 Jul 23&10039 &00:16& S11E65 & X4.8 &Halo &2285& $\cdot\cdot\cdot$& $\cdot\cdot\cdot$\\
2002 Jul 26 & 10046 &15:37& N10W24 & Optical  &$\cdot\cdot\cdot$& $\cdot\cdot\cdot$ &$\alpha$& MFE \\
2002 Jul 26 & 10044 &17:58& S25E20 & C9.5 & $\cdot\cdot\cdot$ &$\cdot\cdot\cdot$&$\beta$& MFE\\

\hline
\end{tabular}
\end{center}
\end{table*}

\clearpage
\begin{table*}{}
\begin{center}
TABLE 2 \\
Continue \\
\vspace{0.5cm}
\begin{tabular}{llllllllll}
\hline\hline \label{tbl-1}
Date & AR & Time & Location  & Flares& CMEs &CME Speed (km $s^{-1}$)&Hale &   MFE or MFC\\
\hline

2002 Aug 14 & 10067 &19:34& N10E17 & M1.4 & CME&323.6 &$\beta$& MFE \\
2002 Aug 18 & 10083&16:41& S19E52 & C6.0 & $\cdot\cdot\cdot$&$\cdot\cdot\cdot$ &$\cdot\cdot\cdot$&$\cdot\cdot\cdot$\\
2002 Aug 19 & 10069 &20:56& S13E33 &M3.1 & CME &804.9& $\beta$$\gamma$$\delta$ &MFE \\
2002 Sep 14 & 10105 &17:07& S07W08 & C4.5 & $\cdot\cdot\cdot$ &$\cdot\cdot\cdot$&$\beta$$\gamma$$\delta$& MFC \\
2002 Sep 18 & 10119 & 23:04& S22E03 & C1.0 & $\cdot\cdot\cdot$ &$\cdot\cdot\cdot$&$\beta$& MFE\\
2002 Sep 21 & 10119 &16:50& S15W33 & C2.6 & CME &431.4 &$\beta$$\gamma$$\delta$& MFE \\
2002 Sep 21 & 10123 &20:11& S11W22 & C2.6  & $\cdot\cdot\cdot$ &$\cdot\cdot\cdot$&$\beta$& MFE\\
2002 Sep 29 & 10134 &06:33& N11E17 & M2.6 & $\cdot\cdot\cdot$ &$\cdot\cdot\cdot$& $\beta$$\gamma$$\delta$&MFC \\
2002 Oct 24 & 10162 &18:04& N23W15 &C7.4 &CME  &311.7& $\beta$$\gamma$$\delta$&MFE \\
2002 Oct 29 & 10162 &02:52& N28W55 & M1.1 & $\cdot\cdot\cdot$ &$\cdot\cdot\cdot$& $\cdot\cdot\cdot$  &$\cdot\cdot\cdot$\\
2002 Oct 31 & 10162 &17:24& N29W90 & C8.0 & CME &160.5& $\cdot\cdot\cdot$  &$\cdot\cdot\cdot$\\
2002 Nov 5 & 10177 &20:51& N14W11 & C3.4  & $\cdot\cdot\cdot$&$\cdot\cdot\cdot$ & $\beta$$\gamma$  &MFC\\
2003 Jan 25 & 10268 &18:32& N10W29 & C4.4 & $\cdot\cdot\cdot$ &$\cdot\cdot\cdot$&$\beta$$\delta$  &MFC\\
2003 Jan 29 & 10266 &20:13& S23W61&C1.1 & $\cdot\cdot\cdot$ &$\cdot\cdot\cdot$&$\cdot\cdot\cdot$  &$\cdot\cdot\cdot$\\
2003 Mar 26 & 10321 &16:57&N08E59&  C2.2 & $\cdot\cdot\cdot$ &$\cdot\cdot\cdot$&$\cdot\cdot\cdot$& $\cdot\cdot\cdot$\\
2003 Mar 29 &10318 &18:27& S11W21 &C8.2 &CME &633.8& $\beta$& MFC\\
2003 May 2 & 10345 &02:25& S18W25 & M1.0 & $\cdot\cdot\cdot$&$\cdot\cdot\cdot$ & $\beta$&MFC\\
2003 May 27 & 10365 & 22:48& S06W14 & X1.3 &Halo &964.5 & $\beta$&MFE\\
2003 May 29 & 10368 & 17:29& S35W14 & M2.8 & CME &266.2&  $\beta$&MFC\\
2003 Jun 2 & 10375 &17:25& N13E70 & M1.8 & $\cdot\cdot\cdot$&$\cdot\cdot\cdot$ & $\cdot\cdot\cdot$&$\cdot\cdot\cdot$\\
2003 Jun 11 & 10375 &20:20& N13W59 & X1.6 & No data &$\cdot\cdot\cdot$&$\cdot\cdot\cdot$  &$\cdot\cdot\cdot$\\
2003 Jun 12 & 10375 &01:06&N16W55& M7.3 & No data &$\cdot\cdot\cdot$& $\cdot\cdot\cdot$ &$\cdot\cdot\cdot$\\
2003 Jun 12 & 10375 &17:15& N13W81&M1.1 &  No data &$\cdot\cdot\cdot$& $\cdot\cdot\cdot$ &$\cdot\cdot\cdot$\\
2003 Aug 25 & 10442 &01:34& S09E38 & C3.6 & CME&574.8 & $\beta$&MFC\\
2003 Sep 29 & 10464 &19:45& N04W41 & C3.8 & $\cdot\cdot\cdot$&$\cdot\cdot\cdot$ &  $\beta$$\gamma$&MFC\\
2003 Oct 24 & 10484 &21:33& N04W05 &M1.0 & $\cdot\cdot\cdot$ &$\cdot\cdot\cdot$& $\beta$$\gamma$$\delta$ &MFC\\
2003 Dec 2 & 10508 &09:24& S19W90 & C7.2 & CME &1392.7& $\cdot\cdot\cdot$&$\cdot\cdot\cdot$\\
2004 Mar 24 & 10582 &18:38& N15E90 & C1.5 & $\cdot\cdot\cdot$&$\cdot\cdot\cdot$ &$\cdot\cdot\cdot$ & $\cdot\cdot\cdot$\\
2004 Mar 24 & 10582 &20:14& N15E90 & C7.4& $\cdot\cdot\cdot$ &$\cdot\cdot\cdot$&$\cdot\cdot\cdot$ & $\cdot\cdot\cdot$\\
2004 Mar 24 & 10582 &23:20& N15E90 & M1.5& $\cdot\cdot\cdot$&$\cdot\cdot\cdot$&$\cdot\cdot\cdot$ & $\cdot\cdot\cdot$\\
2004 Mar 29 & 10582&18:27& N15E16 &C3.4 & $\cdot\cdot\cdot$&$\cdot\cdot\cdot$& $\cdot\cdot\cdot$&No data\\
2004 Mar 30 & 10581 &22:46& S01E04 & C2.0 & No data&$\cdot\cdot\cdot$&$\alpha$&No\\
2004 May 12 & 10609 &15:34& N01E41 & C1.3 & $\cdot\cdot\cdot$&$\cdot\cdot\cdot$& $\beta$&MFE\\
2004 May 21 & 10615 &22:10& N15E02 & C1.2 & $\cdot\cdot\cdot$&$\cdot\cdot\cdot$ & $\alpha$&MFE\\
2004 May 24 & 10615 &22:32& N20W35 & C1.0 & $\cdot\cdot\cdot$&$\cdot\cdot\cdot$& $\beta$&MFE\\
2004 Jul 13 & 10646 &00:00& N14W41 & M6.7 & CME &408.9 &  $\beta$&MFE\\
2004 Jul 26 &10652 &16:52& N16W50 & M1.1 & CME &400.9& $\cdot\cdot\cdot$ &No data\\
2004 Jul 26 &10652&17:31& N03W45&$\cdot\cdot\cdot$&$\cdot\cdot\cdot$&$\cdot\cdot\cdot$&$\cdot\cdot\cdot$ &No data\\
2004 Aug 12 & 10656 &15:58& S15W05 & C6.5 & $\cdot\cdot\cdot$&$\cdot\cdot\cdot$& $\beta$$\gamma$$\delta$ &MFE\\
2004 Aug 18 & 10656 &17:37& S16W90 & X1.8 & CME &601.5& $\cdot\cdot\cdot$&$\cdot\cdot\cdot$\\
2004 Nov 3 & 10696 & 17:54& N05E23 & M1.0 & CME &513.2& $\beta$&MFC\\
2004 Nov 4 & 10696 &21:32&N07E12& M2.5 & CME &1054.5&$\beta$$\gamma$&MFC\\
2004 Nov 4 & 10696 &22:41&N09E09& M5.4 &$\cdot\cdot\cdot$&$\cdot\cdot\cdot$ &$\beta$$\gamma$&MFC\\
2004 Nov 10 & 10696 &02:01& N09W50 & X2.5 & Halo &3387.1& $\beta$$\gamma$$\delta$& MFE\\
2004 Nov 18 & 10700 &19:48& N04W90 & C1.9 & $\cdot\cdot\cdot$&$\cdot\cdot\cdot$&$\cdot\cdot\cdot$&$\cdot\cdot\cdot$\\
2005 Jan 16 & 10720 &07:11& N18W10& C4.4 & $\cdot\cdot\cdot$&$\cdot\cdot\cdot$&$\beta$$\delta$& MFE\\
2005 May 11 &10758 &19:19&S13W52 & M1.1 & Halo  &550.3&$\cdot\cdot\cdot$&$\cdot\cdot\cdot$\\
2005 May 26 & 10767 &20:59& S07E12 & C8.6 & CME &419.9&$\beta$$\gamma$&MFE\\
2005 Jun 1 & 10772 &17:32& S17E41 & C3.1 & $\cdot\cdot\cdot$&$\cdot\cdot\cdot$&$\beta$&MFE\\
2005 Jun 1 & 10772 &21:56&S16E39 & C7.2  & $\cdot\cdot\cdot$&$\cdot\cdot\cdot$&$\beta$&MFE\\
2005 Jun 6 & 10772 &18:19&S16W22 & C1.2 & $\cdot\cdot\cdot$&$\cdot\cdot\cdot$ & $\beta$&MFE\\
2005 Jun 16 & 10775 &19:54& N09W90 & M4.0 & No data &$\cdot\cdot\cdot$& $\cdot\cdot\cdot$&$\cdot\cdot\cdot$\\
2005 Jul 13 &10786 &19:04 & N09W85 & M1.2 &  $\cdot\cdot\cdot$&$\cdot\cdot\cdot$&$\cdot\cdot\cdot$ &$\cdot\cdot\cdot$\\
2005 Jul 27 & 10792 &04:40& N11E90 & M3.7 & Halo &1787.5& $\cdot\cdot\cdot$&$\cdot\cdot\cdot$\\
2005 Jul 28 & 10792 &21:54 & N12E90& M4.8 & CME &1478.4& $\cdot\cdot\cdot$&$\cdot\cdot\cdot$\\
2005 Jul 29 & 10792 &17:39 &N18E71 & C3.4 & CME &296.7& $\cdot\cdot\cdot$&$\cdot\cdot\cdot$\\
2005 Jul 30 & 10792 &05:03& N10E76 & C9.4 & CME &No data&$\cdot\cdot\cdot$ &$\cdot\cdot\cdot$ \\
2005 Jul 30 & 10792 &05:42&N13E70 & X1.3 & Halo &1968.4&$\cdot\cdot\cdot$ & $\cdot\cdot\cdot$\\
2005 Sep 13 & 10808 &18:00& S11E11 & X1.5 & Halo &1866.1& $\beta$$\gamma$$\delta$ &MFE\\
2006 Jul 6 & 10898 &08:06& S10W38 & M2.5 & Halo &910.6& $\beta$&MFC\\
2007 Mar 2 & 10944 &04:47& S01W17 & B2.5 & No data&$\cdot\cdot\cdot$&$\alpha$&MFC\\

\hline
\end{tabular}
\end{center}
\end{table*}

\end{document}